\documentclass[]{spie}  
\pdfoutput=1

\usepackage{amsmath,amsfonts,amssymb}
\usepackage{graphicx}
\usepackage[colorlinks=true, allcolors=blue]{hyperref}

\title{Spatial polarization modulators: distinguishing diffraction effects from spatial polarization modulation}

\author[a,b]{Willeke Mulder}
\author[a]{David S. Doelman}
\author[a]{Christoph U. Keller}
\author[c]{C.H. Lucas Patty}
\author[a]{Frans Snik}
\affil[a]{Leiden Observatory, Leiden University, PO Box 9513, 2300 RA Leiden, The Netherlands;}
\affil[b]{Faculty of Aerospace Engineering, Delft University of Technology, Kluyverweg 1, 2629 HS Delft,The Netherlands;}
\affil[c]{Physikalisches Institut, Universitat Bern, Sidlerstrasse 5, 3012 Bern, Switzerland;}

\authorinfo{Further author information: (Send correspondence to W.M.)\\W.M.: E-mail: wmulder@strw.leidenuniv.nl, home.strw.leidenuniv.nl/~wmulder/}

\begin{document} 
\maketitle

\begin{abstract} 
Are we alone? In our quest to find life beyond Earth, we use our own planet to develop and verify new methods and techniques to remotely detect life. Our Life Signature Detection polarimeter (LSDpol), a snapshot full-Stokes spectropolarimeter to be deployed in the field and in space, looks for signals of life on Earth by sensing the linear and circular polarization states of reflected light. Examples of these biosignatures are linear polarization resulting from O2-A band and vegetation, e.g. the Red edge and the Green bump, as well as circular polarization resulting from the homochirality of biotic molecules. LSDpol is optimized for sensing circular polarization. To this end, LSDpol employs a spatial light modulator in the entrance slit of the spectrograph, a liquid-crystal quarter-wave retarder where the fast axis rotates as a function of slit position. The original design of LSDpol implemented a dual-beam spectropolarimeter by combining a quarter-wave plate with a polarization grating. Unfortunately, this design causes significant linear-to-circular cross-talk. In addition, it revealed spurious polarization modulation effects. Here, we present numerical simulations that illustrate how Fresnel diffraction effects can create these spurious modulations.
We verified the simulations with accurate polarization state measurements in the lab using 100\% linearly and circularly polarized light.

\end{abstract}


\section{INTRODUCTION}
\label{sec:intro}  




Homochirality is an unambiguous bio-signature that can be detected using polarimetry in the visible  spectrum \cite{Gil14}\cite{Patty17}\cite{Patty19}\cite{Patty21} . Life as known on Earth can therefore be detected by looking at the circular polarization of reflected sunlight. A polarimeter in orbit around the Earth could provide an observational assessment of the average circular polarization signal of planet Earth, which is the relevant quantity to assess the potential to see signs of homochirality on an unresolved exoplanet. The optical requirements of LSDPol are based on the ultimate goal of a full-Stokes spectropolarimeter in space; hence the optical design should be compact, lightweight and avoid moving parts.\cite{Snik19} 

LSDPol measures the polarization of scattered sunlight as a function of wavelength. The LSDpol prototype design prioritizes accurate and sensitive polarimetry over spectroscopy and angular resolution. Therefore, the very first element is a patterned liquid crystal retarder\cite{Snik19}\cite{Keller20} that modulates the polarization of the incoming light. Sensitive polarimeters commonly measure two orthogonal polarization states simultaneously to minimize artefacts such as those due to variations in the spectrograph slit illumination. LSDpol combined polarimetry and spectroscopy with a fixed quarter-wave retarder and a polarization grating to form a linear polarizing beam splitter that allows for a measurement of two orthogonal polarization states over the visible spectrum in a single snapshot\cite{Snik19}\cite{Keller20} .

The zeroth order of the polarization grating re-images a homogeneously illuminated entrance slit. This image should, in theory, be homogeneous as well. We can detect possible inhomogeneous illumination of the entrance slit by inspecting the projected 0th order. In the lab, Keller et al. (2020)\cite{Keller20} found that this slit image is modulated even in the case of incoming unpolarized light and that the spectra of contain a spurious modulation with a frequency equal to a V modulation, but phase-shifted by 90 degrees as compared to the modulation for incoming circularly polarized light. These will be referred to as the \textit{spurious signal} and the \textit{shifted V lookalike}, respectively. A simple Mueller matrix model, or short \textit{Mueller model}, can include deviations from perfect retardance or fast axis orientation angles. However, it can not reproduce diffraction effects, and it could not explain the spurious features. Therefore, we decided to create a realistic diffraction simulation to investigate the effect of both Fresnel and Fraunhofer diffraction on the detector image. 

This proceeding has the following structure. In Section \ref{sec:theory} we explain the working principle and the prototype design of LSDPol and summarize the discrepancies between theory and lab measurements that were found during the initial data analysis. 
We numerically investigate the source of the spurious signal using simulations that include Fresnel diffraction. Thereafter, the results of the simulations were verified with additional measurements in the lab. The results of the simulation and measurements will follow respectively in Section \ref{sec:simulations} and \ref{sec:measdistance} after which we will discuss our main results and conclusions in Section \ref{sec:conclusions}.



\section{Full-stokes spectropolarimeter}
\label{sec:theory}  

Sparks et al.\ (2012)\cite{Sparks12} combined classical spectroscopy along one direction and polarization as an intensity modulation in the orthogonal direction into a snapshot full-Stokes spectropolarimeter. Their approach is based on a spatially variable retarder. Sparks et. al (2019) used the concept of a quarter wave plate whose fast axis direction changes with location to demonstrate the concept of a full-stokes spectropolarimeter. This principle is described in Section \ref{sec:polarimetry} and Section \ref{sec:spectrometry}. 

The concept of Sparks et. al (2019) include moving parts and will therefore not meet the instrument requirements of a compact instrument with no moving parts\cite{Snik19}. This is why LSDpol is based on liquid crystal direct-write technology to create a patterned close-to-quarter-wave retarder. In Section \ref{sec:design} we discuss the main components of the LSDPol prototype.

\subsection{Polarimetry}
\label{sec:polarimetry}
At least two intensity measurements are required for a polarization measurement \cite{Berry1977MeasurementOT}. Polarimeters can use both the principle of temporal or spatial modulation. An example of a temporal modulation is the combination of a physically rotating retarder in front of a linear polarizer. The modulated intensity is a function of the orientation of the fast axis of the retarder $\theta$ and the retardance $\delta$ of the retarder. The intensity modulation introduced by a perfect rotating retarder followed by an ideal linear polarizer is given by 
\begin{equation} \label{eqn:mod}
I^{\prime}(\theta)=\frac{1}{2}\left(I+\frac{Q}{2}((1+\cos \delta)+(1-\cos \delta) \cos 4 \theta)+\frac{U}{2}(1-\cos \delta) \sin 4 \theta-V \sin \delta \sin 2 \theta\right),
\end{equation} where $I^{\prime}$ is the measured intensity as a function of the fast-axis orientation angle $\theta$, $\delta$ is the retardance of the retarder and $I$, $Q$, $U$ and $V$ are the Stokes parameters of the incoming light. 

\begin{figure} [ht]
   \begin{center}
   \begin{tabular}{c} 
   \includegraphics[height=5cm]{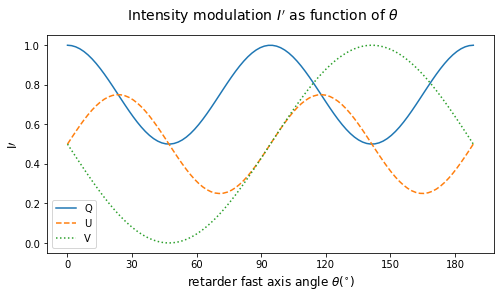}
	\end{tabular}
	\end{center}
   \caption[example] 
   { \label{fig:modulation} The modulated observed intensity ($I^\prime$) of fully linearly ($Q=I$, $U=I$) or circularly ($V=I$) polarized light after passing through a rotating quarter-wave retarder with a fast axis orientation of $\theta$.
}
   \end{figure}

Equation \ref{eqn:mod} implies maximum circular polarization sensitivity for an ideal quarter-wave retarder with $\delta=\frac{\pi}{2}$ when the fast-axis orientation angle covers $0<\theta<\pi$. Figure \ref{fig:modulation} shows the resulting intensity modulation $I^{\prime}(\theta)$. The blue, orange and green lines represent the intensity modulations of fully-polarized incoming light, respectively $Q=I$, $U=I$ and $V=I$. Circular Stokes $V$ exhibits a 100\% modulation, while linear Stokes $Q$ and $U$ only exhibit 50\% modulation.
The lower sensitivity to linear polarization is one of the design trade-offs that were made to prioritize the measurement of \textit{Stokes-V}. 

\subsection{Spectroscopy}
\label{sec:spectrometry}
\begin{figure} [ht]
   \begin{center}
   \begin{tabular}{c} 
   \includegraphics[width=0.95\textwidth]{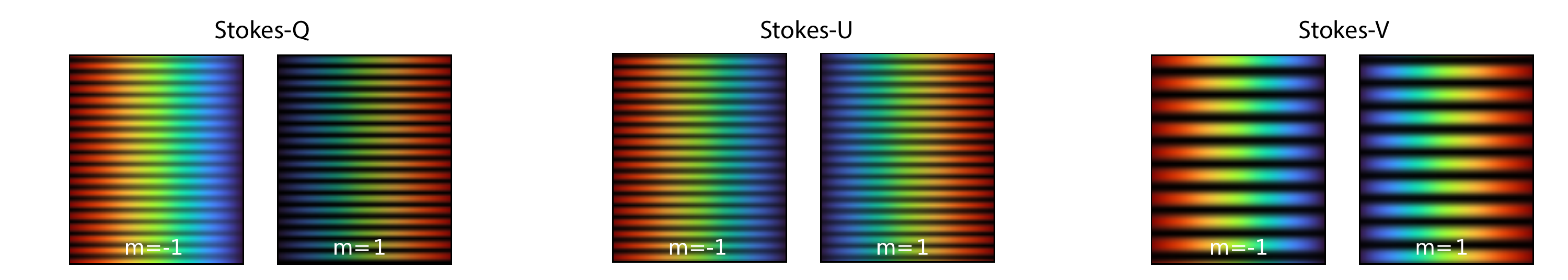} 
	\end{tabular}
	\end{center}
   \caption[example] 
   { \label{fig:muellerQUV} From left to right, the simulated responses of the first orders of the polarization gratings for fully polarized light with Stokes $\pm$Q=I, $\pm$U=I, $\pm$V=I, see Equation\ref{eqn:mod}. The colors indicate the spectrum and the brightness shows the intensity modulation. Note that this Mueller model does not include diffraction effects, except for the dispersion of the polarization grating. 
}
   \end{figure}

Light can be dispersed with an ordinary diffraction grating. This allows us to use one dimension of an imaging detector for spectroscopy and the other dimension for polarimetry. These dimensions will be referred to as the \textit{spectral and polarimetric dimensions}.
By replacing the ordinary grating with a polarization grating\cite{Oh08}\cite{Packham10}, light in the same diffraction order but with the opposite sign corresponds to orthogonal circular polarization states. Thereby, the polarization grating also acts as a circular polarization beam splitter. We can transform the circular-polarization beam splitter into a linear-polarization beam splitter by placing an additional quarter-wave retarder in front of the grating\cite{Chen19}. Figure \ref{fig:muellerQUV} shows the response of the polarization modulator, the quarter-wave retarder and polarization grating to fully polarized light. The images reveal the response to $\pm$Q=I, $\pm$U=I, $\pm$V=I polarized light, respectively. The horizontal axis represents the spectral dimension and the vertical axis represent the polarimetric dimension. The rainbow colors indicate the diffraction of light in the visible wavelength regime. The patterned quarter-wave retarder is responsible for the intensity modulation pattern shown as the brighter and darker areas.

\subsection{Initial prototype design}
\label{sec:design}
To measure all Stokes parameters, LSDPol combines the ides behind temporal and spatial polarization modulation. The liquid-crystal based patterned quarter-wave retarder mimics a rotating quarter-wave plate by rotating the fast-axis orientation angle, $\theta$, linearly along the vertical spatial dimension. We will refer to this element as the \textit{polarization modulator} or just \textit{modulator}. Since we did not know in advance what rotation speed would be optimal, the patterned retarder increases the frequency of rotation increases 10 times by a factor of 2 along the horizontal axis. We will refer to this frequency as the \textit{modulation frequency}. 
Having multiple frequencies patterned on our modulator allows us to select the optimum frequency. 

\noindent The optical design can be described as follows:
\begin{itemize}
    \item Optics can introduce small amounts of cross-talk from linear to circular polarization or may even induce circular polarization from unpolarized light. Therefore, the incoming light is first modulated by the polarization modulator right as it enters the instrument. 
    \item The polarization modulator also acts as an inefficient polarization grating, and it will therefore disperse some of the incoming circularly polarized light into left and right-handed circular components. The three custom-made baffles in front of the modulator select a specific field of view. This ensures that left- and right-handed circular polarization comes from exactly the same field of view. 
    \item The modulator is collocated with a 0.1 by 10 mm entrance slit that directly follows it. In this design, the modulator can be displaced perpendicular to the slit to select the desired modulation frequency.
    \item A collimating lens minimizes the divergence of the beam on the polarization grating to maximize the spectral resolution for a chosen field of view.
    \item An internal field stop after the lens can also be used to define the field of view.
    \item The modulation is analyzed by a fixed quarter-wave retarder, QWR, with its fast axis at $45^{\circ}$ with respect to the rules of the polarization grating.  The QWR transforms the polarization states of the modulated light from linear to circular and vice versa.
    \item A $10$-degree polarization grating, PG, diffracts the incident light into two beams that have opposite circular polarization. In combination, the QWR and PG act as a linear polarizing beam splitter. LSDPol uses the combination of the polarization modulator in combination with the linear polarizing beam splitter to image two individual, orthogonal polarization states in one image (snapshot).
    \item The last optical element is a camera lens that creates an image of the dispersed slit on the detector. 
\end{itemize}

\subsection{Discrepancies between observations and theory}
Combining a polarization modulator with a polarizing beam splitter is an elegant way to perform snapshot full-Stokes spectropolarimetry. However, the current prototype suffers from several issues. 
\begin{itemize}
    \item The instrument introduces linear to circular crosstalk because any retardance error of the fixed quarter-wave retarder causes a wavelength-dependent polarization cross-talk. Deviations from the fast axis orientation only cause a change in the absolute phase of the modulation, which is of no consequence. We could determine the retardation of the quarter-wave retarder in front of the polarization grating, but this approach has not yet been successful. 
    \item Linear-polarizer calibration images of the original LSDPol design reveal a spurious signal in the zeroth order diffraction, and potentially even more troublesome, a \textit{Stokes-V}-like modulation. This could be due to the polarization modulator acting as a weak grating. Only a full diffraction simulation can reveal the impact of the polarization modulator acting as a weak grating. 
\end{itemize}

The latter is the main purpose of this proceeding. The impact of diffraction effects is investigated using both simulations and measurements in the lab. Two main set-ups are simulated and used to fully decouple the effect of the quarter-wave retarder and polarization grating from the polarization modulator. 

The original prototype contains a quarter-wave retarder combined with a polarization grating. This set-up will be referred to as \text{LSDPol$_{\text{QWR}}$}. To disentangle the impact of the Fresnel propagation from the imperfect quarter-wave retardance, another set-up replaces the retarder with a wire grid polarizer. This set-up is referred to as \text{LSDPol$_{\text{WGP}}$}. The results of the latter is not further discussed in this proceeding.

In both set-ups we replaced the 10-degree polarization grating with a 5-degree polarization grating. The latter has a smaller dispersion. This allows us to trade off a lower wavelength resolution for reduced barrel distortion, which is most prominent at the edges of the detector. The distortion leads to an apparent tilt in the modulation intensity, which changes the frequency of the modulation along the pixel columns. This can be corrected for once we fully understand the LSDPol prototype.

\section{Numerical simulation including Fresnel diffraction}
\label{sec:simulations}
The Mueller-matrix model can describe the polarization modulation of a rotating quarter-wave retarder.\cite{Sparks19} It can also describe the patterned quarter-wave retarder that rotates its fast-axis along the vertical spatial dimension.\cite{Snik19}\cite{Keller20}. However, the incoherent Mueller model cannot reproduce potential diffraction effects, which causes changes in the amplitude and phase of the modulated intensity as a function of distance from the polarization modulator. These could be falsely identified as polarization signals or instrumental cross-talk.  

Raw linear-polarizer calibration images of LSDPol reveal a spurious signal in the zeroth diffraction order along the polarimetry dimension that was not expected by the Mueller model.\cite{Keller20} The liquid crystal spatial polarization modulator acts as a weak polarization grating. A diffraction that is created in the first optical element might explain why we see both the spurious signal in the zeroth order and the shifted V lookalike in the dispersed spectra. The first step to identify the cause of these features consists of a coherent  simulation that includes both diffraction and polarization effects.

With our diffraction simulation, we want to simulate the response to polarized light and compare the results to the Mueller model. Our hypothesis is that Fresnel diffraction effects are causing the spurious signal that was observed. The results of our numerical simulation are revealed in Section \ref{sec:sim_fullpol}. 
The next step involves increasing the distance between the polarization modulator and the entrance slit. This distance will be referred to as the \textit{propagation distance}.  Section \ref{sec:sim_0thFresnel} describes the effect of the signal in the zeroth order while varying the propagation distance. Section \ref{sec:sim_2stFresnel} reveals the different responses in the $\pm 1$ diffraction orders due to a varying propagation distance. Lastly, the simulated response to unpolarized light reveals a possible explanation for the shifted V lookalike. This is illustrated in Section \ref{sec:sim_unpol}. 

These simulations provide us with a better understanding of the addition effects of Fresnel diffraction in the LSDPol prototype and whether Fresnel diffraction can explain the spurious signal and shifted V lookalike.

\subsection{Fresnel propagation of a monochromatic wavefront}

Our diffraction simulation is based on HCIPy\cite{Por18}. HCIPy is a Python package that can be used to perform end-to-end simulations of astronomical high-contrast imaging instruments.

\begin{figure} [ht]
  \begin{center}
  \begin{tabular}{c} 
  \includegraphics[width=0.95\textwidth]{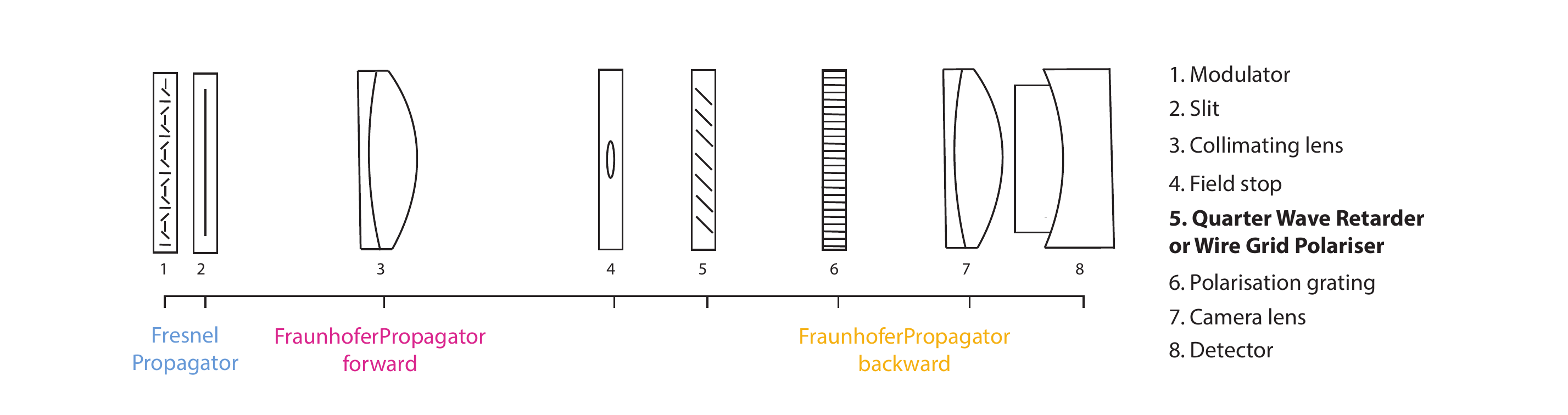}
	\end{tabular}
	\end{center}
  \caption[example] 
  { \label{fig:set-up}  A schematic overview of the optical components that are simulated numerically. The numbers below the components refer to the numbers in front of the descriptions on the right. The corresponding diffraction approximations are noted below the horizontal line. The FresnelPropagator describes the propagation of light over the \textit{propagation distance}. The FraunhoferPropagator forwards is used to simulate the collimating lens and the FraunhoferPropagator backwards to simulate the camera lens.  
}
  \end{figure}

Our simulation covers the entire system optic-by-optic. The simulations reveal how either (i)fully polarized or unpolarized light passes through the (ii) polarization modulator, (iii) entrance slit, (iv) collimating lens, (v) field stop, (vi) quarter-wave retarder, (vii) polarization grating and is imaged through a camera lens onto a detector plane. A sketch of the set-up is provided in Figure \ref{fig:set-up}. 

The following calculations are made for each individual wavelength:
\begin{itemize}
    \item[(i)] The input of the system is an incoming, flat wavefront with a size that slightly exceeds the height of the entrance slit. The wavefront is monochromatic and can be fully polarized.  
    \item[(ii)] The wavefront passes through the liquid crystal polarization modulator. The modulator is represented by a linear retarder with a retardation of $\delta=\pi/2$ that we rotate by defining a separate, rotating phase ramp along the vertical axis. 
    \item[(iii)] The wavefront propagates through the modulator, after which we expect diffraction effects to occur. Therefore, Fresnel propagation is used to propagate a small distance of 3 mm towards the slit. The slit is represented by a rectangular aperture with a height of 10 mm and a width of 0.1 mm. 
     \item[(iv)] The collimating lens is simulated as a forward Fraunhofer propagator from the pupil plane to the focal plane. It transforms the electric field in the slit aperture into the electric field in the field stop.
    \item[(v)] The field stop is a circular aperture with a diameter of 4 mm that is defined in the focal plane where it limits the extent of the transmitted electric field. 
    \item[(vi)] A quarter-wave retarder is also placed in the focal plane. After the retarder, we propagate back to the pupil plane by using a backwards Fraunhofer propagation. 
    \item[(vii)] A polarization grating is the same as our polarization modulator, except for the retardation being half a wave. Here we assume a retardation of $\delta=0.93\pi$ to simulate the leakage into the 0-th order. The electric field in the detector plane then contains the 0-th as well as the $\pm1$-st orders.
\end{itemize}

\subsection{Response to a fully polarized light source}
\label{sec:sim_fullpol}
\begin{figure} [ht]
   \begin{center}
   \begin{tabular}{c} 
   \includegraphics[width=0.30\textwidth]{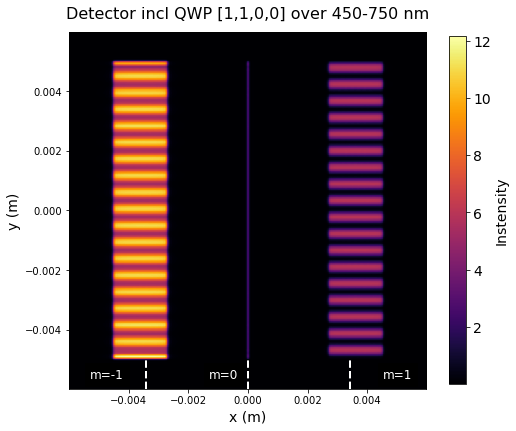} 
   \includegraphics[width=0.30\textwidth]{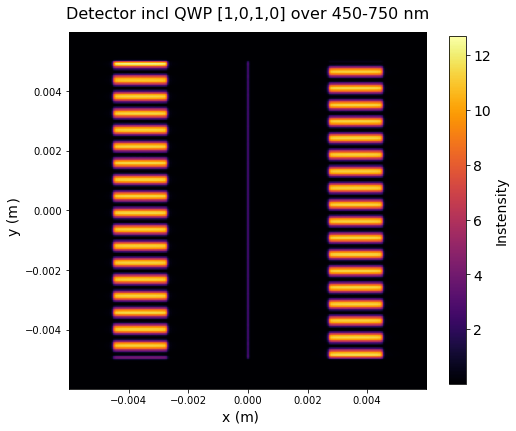} 
   \includegraphics[width=0.30\textwidth]{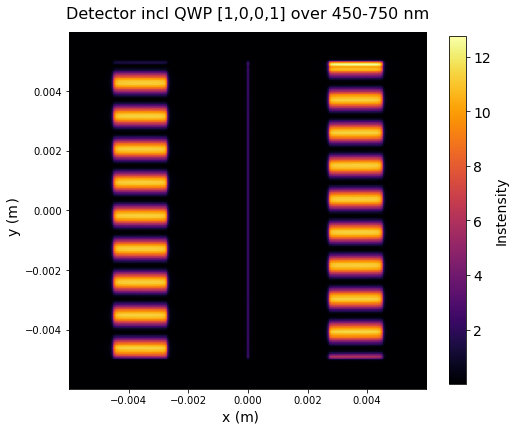} 
	\end{tabular}
	\end{center}
   \caption[example] 
  { \label{fig:simuQUV_qwp} Numerical simulation of LSDPol$_{QWR}$ showing the response to \textit{Q}, \textit{U} and \textit{V}. The white lines indicate the locations of the first (m=$\pm1$) and zeroth (m=0) polarization grating diffraction orders. The propagation distance from the modulator to the slit was set to 0 mm.
}
   \end{figure}

The performance of our simulation can be compared to the Mueller model from Figure \ref{fig:muellerQUV}. This model was created using simple Mueller matrix multiplications of the individual polarimetric elements, leading to Equation \ref{eqn:mod}. All the simulations in this proceeding are carried out for a broader wavelength range from 450 nm to 750 nm with a resolution of 1 nm. The resulting intensities are combined on the \textit{detector plane}.  Figure \ref{fig:simuQUV_qwp} shows the intensity on the detector plane for three different sources that are fully polarized with $Q=I$, $U=I$ or $V=I$. We will refer to these results as \textit{Q}, \textit{U} and \textit{V}.

The images in Figure \ref{fig:simuQUV_qwp} reveal similar intensity modulation patterns as those from the Mueller model. There is the intensity imbalance in Q, which is expected from the additional $\frac{Q}{2}\left(1+\cos\delta\right)$ term in Equation \ref{eqn:mod}. We conclude that our simulation creates the expected intensity modulation for a fully polarized light source.

The next step involves increasing the propagation distance between the polarization modulator and the entrance slit, corresponding to optical elements 1 and 2 in Figure \ref{fig:set-up}. The wavelength range is the same as those of Figure \ref{fig:simuQUV_qwp}. The modulation frequency is increased by a factor of 2. Simulations for \textit{Q} are done while changing the propagation distance from 0 to 60 mm in 5 mm steps. The simulations for 0 mm, 30 mm and 60 mm are compared to observations, and therefore presented in this proceeding.

\subsection{Zeroth grating order variation due to Fresnel propagation}
\label{sec:sim_0thFresnel}
\begin{figure} [ht]
   \begin{center}
   \begin{tabular}{c} 
  \includegraphics[width=0.49\textwidth]{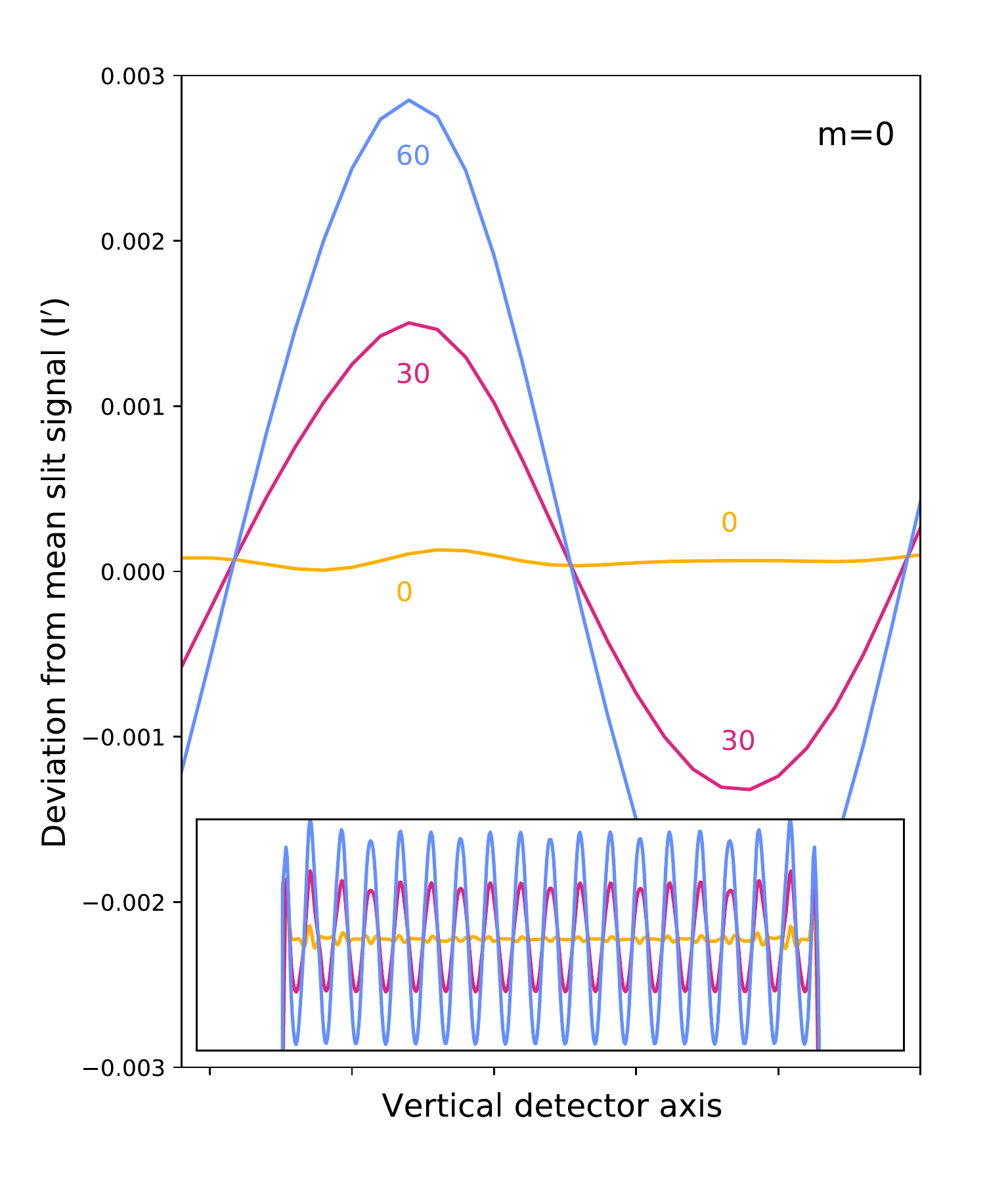}
	\end{tabular}
	\end{center}
   \caption[example] 
   { \label{fig:simu_HCIPy_modulation0th} The yellow, pink and blue lines show the deviation from the expected homogeneous illuminated signal in the zeroth order (m=0) to the simulated \textit{Q} response. The horizontal axis is a zoom-in of the full intensity along the vertical detector plane, represented in the bottom plot. The variation increases as a function of propagation distance. The lower panel reveals a second oscillation of half the modulation frequency on top of the variation.
}
   \end{figure}

The signal in the zeroth diffraction order is obtained from the central pixel column.  We focus on variations along the slit image by subtracting the mean along the vertical axis. The remaining signal is the deviation from the expected homogeneous illuminated signal. The variation in intensity caused by a propagation distance of 0 mm, 30 mm and 60 mm are represented by the yellow, pink and blue lines in Figure \ref{fig:simu_HCIPy_modulation0th}. For the sake of consistency, the same color coding will be used for identical propagation distances in the plots below. 

For a distance of 0 mm, the instrument gives the response that largely coincides with the spatial modulation theory presented in Equation \ref{eqn:mod}. The fluctuation of $\pm0.0001$ are likely due to diffraction at the field stop. The amplitude of the variation grows with increasing propagation distance. The frequency of the variation is equal to the frequency of the intensity modulation. Therefore, the effect will become more prominent for higher modulation frequencies. The blue line also reveals an additional oscillation of about 1/4 of the modulation frequency. Based on these simulations, we conclude that the spurious signal is likely caused by diffraction effects. 


\subsection{First grating order variation due to Fresnel propagation}
\label{sec:sim_2stFresnel}
\begin{figure} [ht]
   \begin{center}
   \begin{tabular}{c} 
  \includegraphics[width=0.95\textwidth]{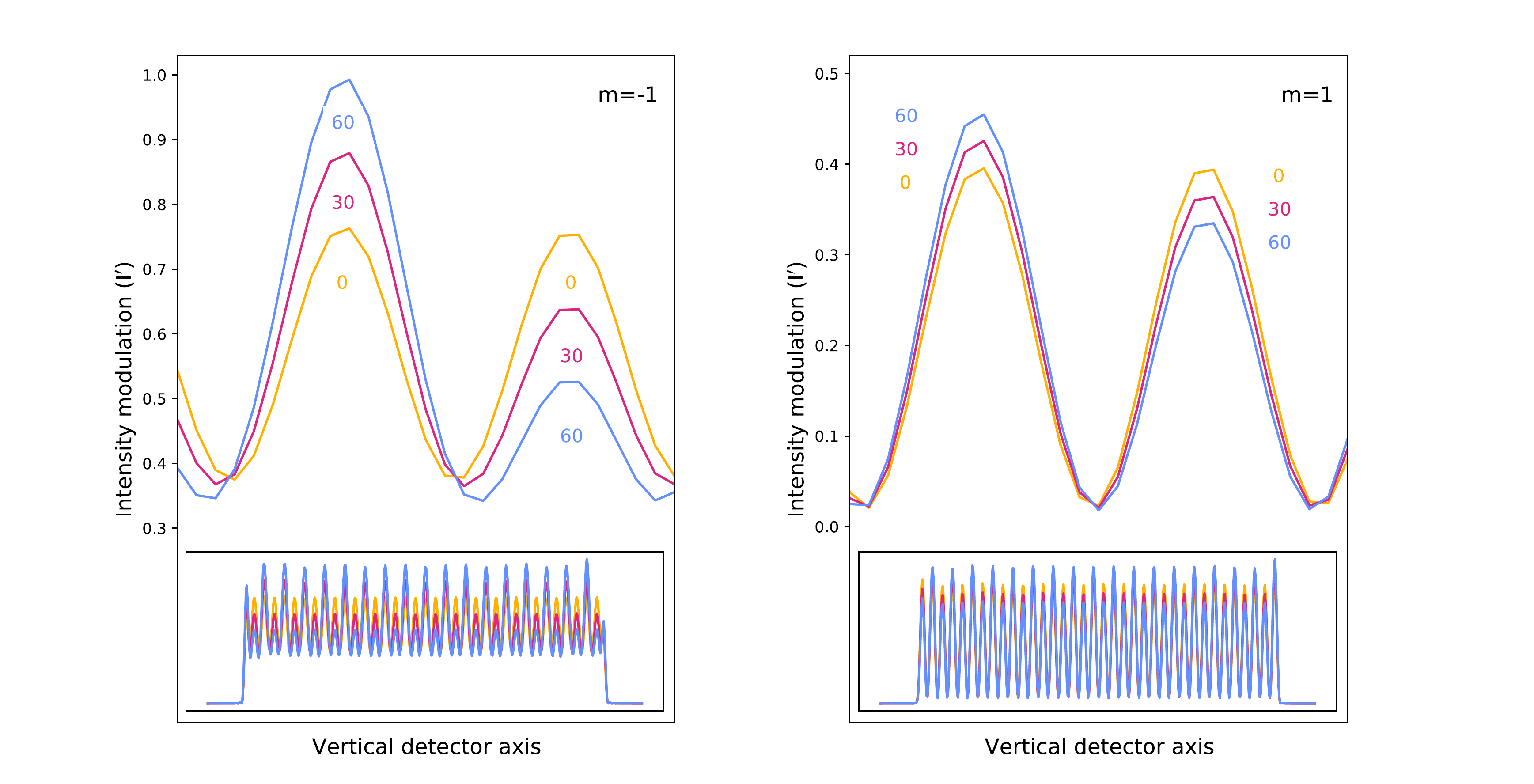}
	\end{tabular}
	\end{center}
   \caption[example] 
   { \label{fig:simu_HCIPy_modulation1st} The yellow, pink and blue lines show the intensity modulation in the first orders (m=-1 on the left, m=0 on the right) to the simulated \textit{Q} response. The horizontal axis is a zoom-in of the full intensity along the vertical detector plane, represented in the bottom plots. The modulation increases as a function of propagation distance. The lower panel reveals the intensity modulation along the slit.
}
   \end{figure}

The polarization grating diffracts the light into the 1st orders with up to 96\% diffraction efficiency. Figure \ref{fig:simuQUV_qwp} indicates the location of diffraction orders (m=$\pm$1). Figure \ref{fig:simu_HCIPy_modulation1st} shows the intensity along the dispersed slit image for fully linearly polarized light in the $+Q$ direction.

The left and right plots show the offset of \textit{Q} due to the additional  $\frac{Q}{2}\left(1+\cos\delta\right)$ term in Equation \ref{eqn:mod}. This difference allows us to quickly distinguish the horizontal and vertical polarized light ($\pm$Q=I). The intensity of the 0-mm simulation is only modulated by a total of 35\% instead of the expected 50\% from Figure \ref{fig:modulation}. While the amplitude of the intensity modulation grows with increasing propagation distances, a secondary modulation at half the frequency occurs. This undesired variation is stronger for m=-1 than it is for m=1. The change in intensity modulation depends on the polarization state of the light source and is a function of the propagation distance. By rotating the fixed quarter-wave retarder, the variation in the intensity modulation also changes. 
The current calibration pipeline of LSDPol uses the intensity of both orders to determine the incoming intensity. It is important to understand these dissimilarities to consider additional diffraction effects in the calibration. 

\subsection{Response to an unpolarized light source}
\label{sec:sim_unpol}
\begin{figure} [ht]
   \begin{center}
   \begin{tabular}{c} 
   \includegraphics[width=0.30\textwidth]{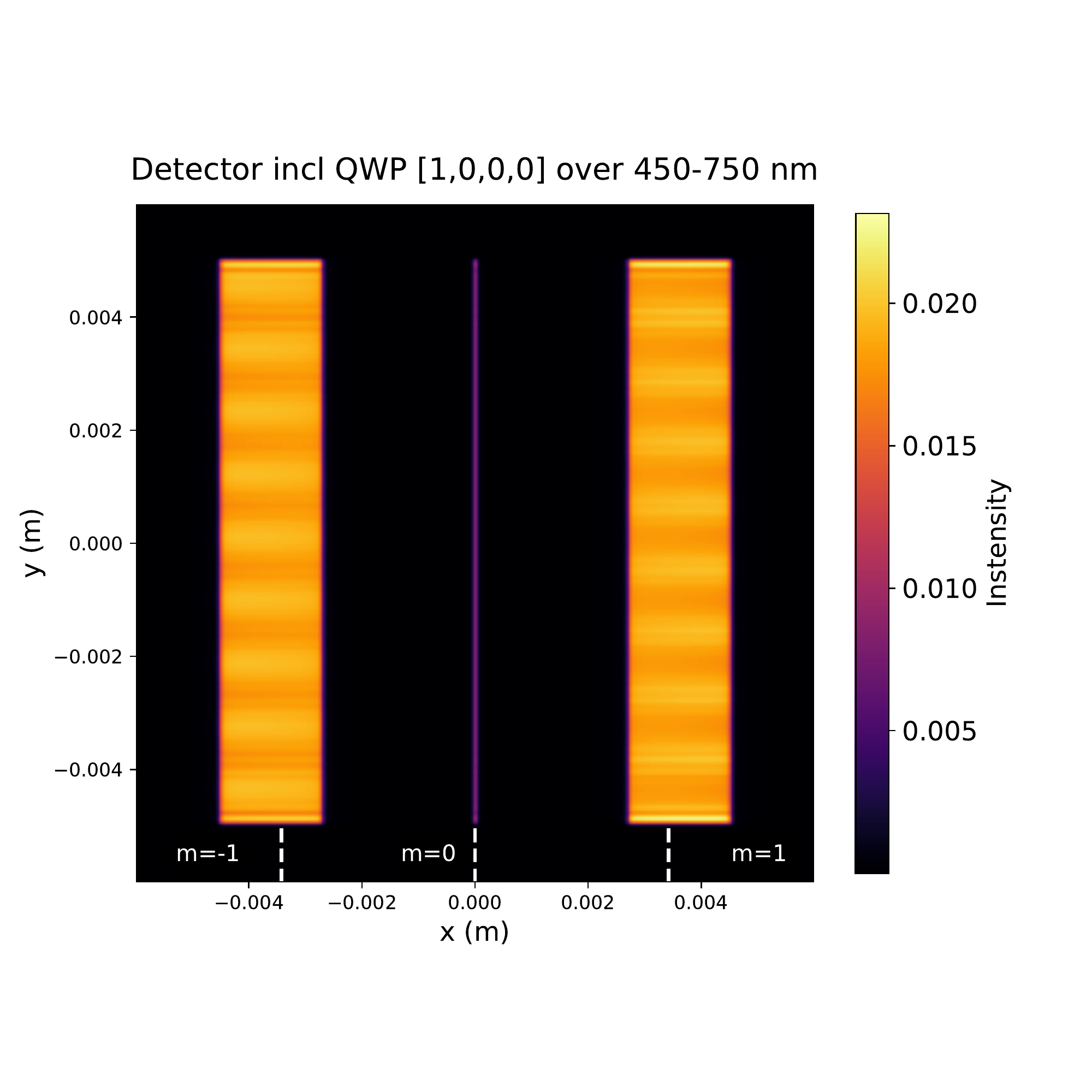}
   \includegraphics[width=0.30\textwidth]{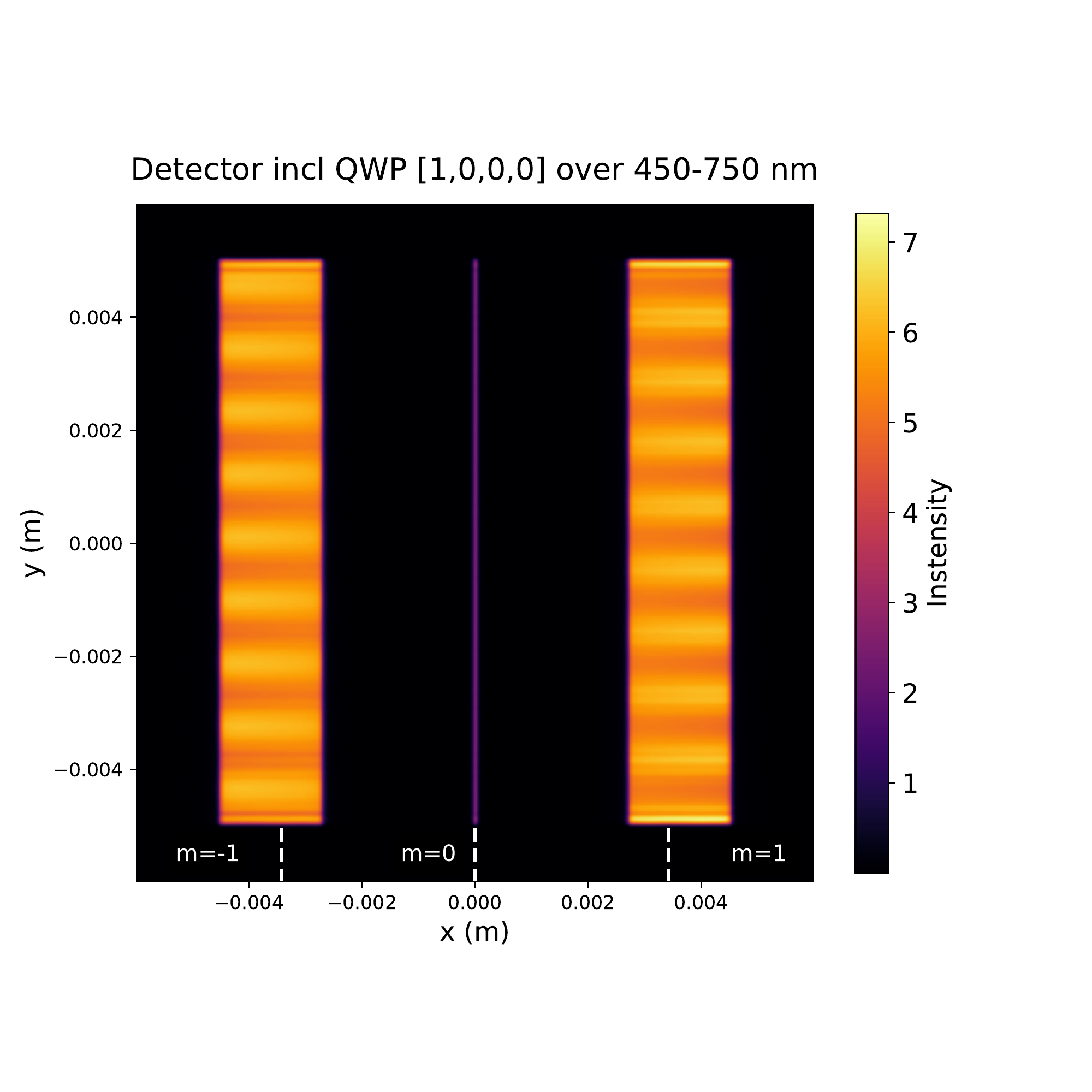}
   \hspace{20pt}
   \includegraphics[width=0.35\textwidth]{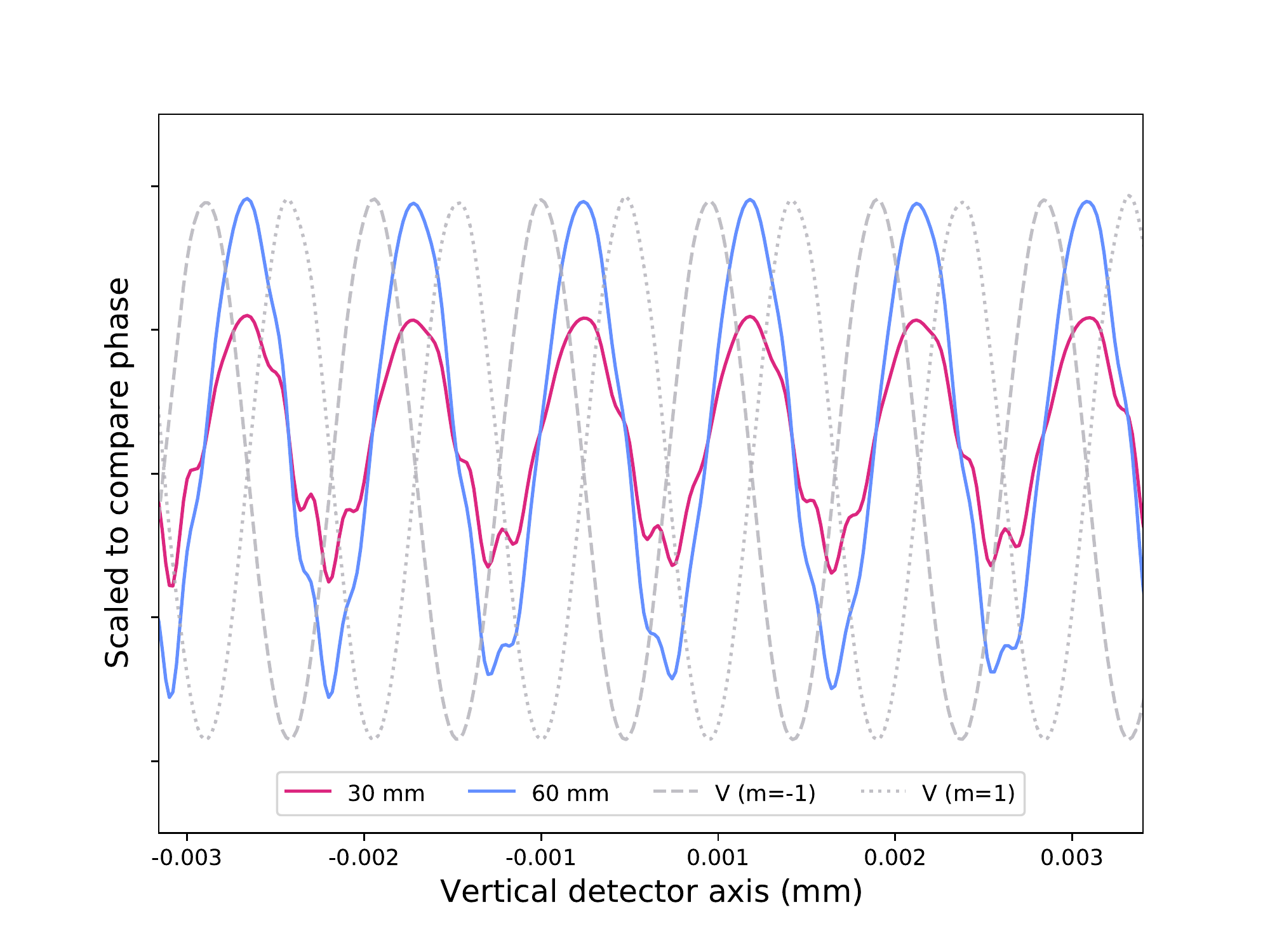}
	\end{tabular}
	\end{center}
   \caption[example] 
  { \label{fig:simuunpolprop_qwp} The left and the middle plot reveal the response to unpolarized light (I=1, Q=U=V=0). The right plot presents the oscillation frequency for 30 and 60 mm from the m=-1 diffraction orders on the left. The dotted lines reveal the corresponding frequencies of the response to \textit{V} for both m=-1 and m=1. The unpolarized light reveals the same frequency as \textit{V} but with a 90$^{\circ}$ phase shift. 
}
   \end{figure}

We gain a better understanding of possible diffraction effects in LSDPol by running simulations for an unpolarized light source. The response to unpolarized light as a function of propagation distance for 30 and 60 mm is presented in Figure \ref{fig:simuunpolprop_qwp}. The response to 0 mm was nearly homogeneous and is therefore not shown. 

The two images on the left reveal the detector plane after imaging an unpolarized light source. The propagation distances were set to 30 and 60 mm, respectively. The right plot compares the signals in the $m=-1$ grating order. The unpolarized light reveals the same frequency as \textit{V} with a 90 $^{\circ}$ phase shift. For a proper comparison to \textit{V} we plot the response to \textit{V} as well with dotted lines. The intensity modulations of the unpolarized light and \textit{V} are scaled to fit in a single plot. The plot can only be used to compare the frequency and phase of the intensity modulations. The unpolarized light exhibits the same modulation frequency as \textit{V} but with a 90 $^{\circ}$ phase shift. 


\section{Measurements using LSDPol prototype set-up}
\label{sec:measdistance}
Our numerical simulations confirm that the observed discrepancies with the Mueller model are (at least partly) caused by diffraction. Here, we present the results of lab measurements of the zeroth and the $\pm 1$ diffraction orders to validate our simulations and confirm that diffraction effects are indeed causing the discrepancies with the Mueller model predictions. 

To do so, we change the physical distance between the polarization modulator and the entrance slit in the LSDPol$_{\text{QWR}}$ setup. As in the simulations, measurements are done for propagation distances ranging from a 0 to 60 mm at 5-mm intervals. The uncertainty in these distances is $\pm1.5$mm because the minimum physical distance between the polarization modulator and entrance slit is $\sim$1.5 $\pm1.5$ mm. 

For each measurement, a flat and dark field correction is made to correct for detector effects. The flat field consists of a detector image while pointing at an unpolarized, diffuse white source that illuminates the detector uniformly. It is crucial for the calibration process to have an accurate flat field. 

In Section \ref{sec:meas_overview} we present dark-corrected detector images of \textit{Q} for distances of  $1.5 \pm 1.5$mm, $31.5 \pm 1.5$mm and $61.5 \pm 1.5$mm. These are referred to as 0 mm, 30 mm and 60 mm respectively. In Section \ref{sec:meas_0th} and Section \ref{sec:meas_1st} we will take a closer look at the effects occurring in the zeroth (m=0) and the two first (m=$\pm$1) diffraction orders. 

\subsection{Change of intensity modulation due to modulator-slit distance }
\label{sec:meas_overview}
\begin{figure} [ht]
   \begin{center}
   \begin{tabular}{c} 
  \includegraphics[width=0.97\textwidth]{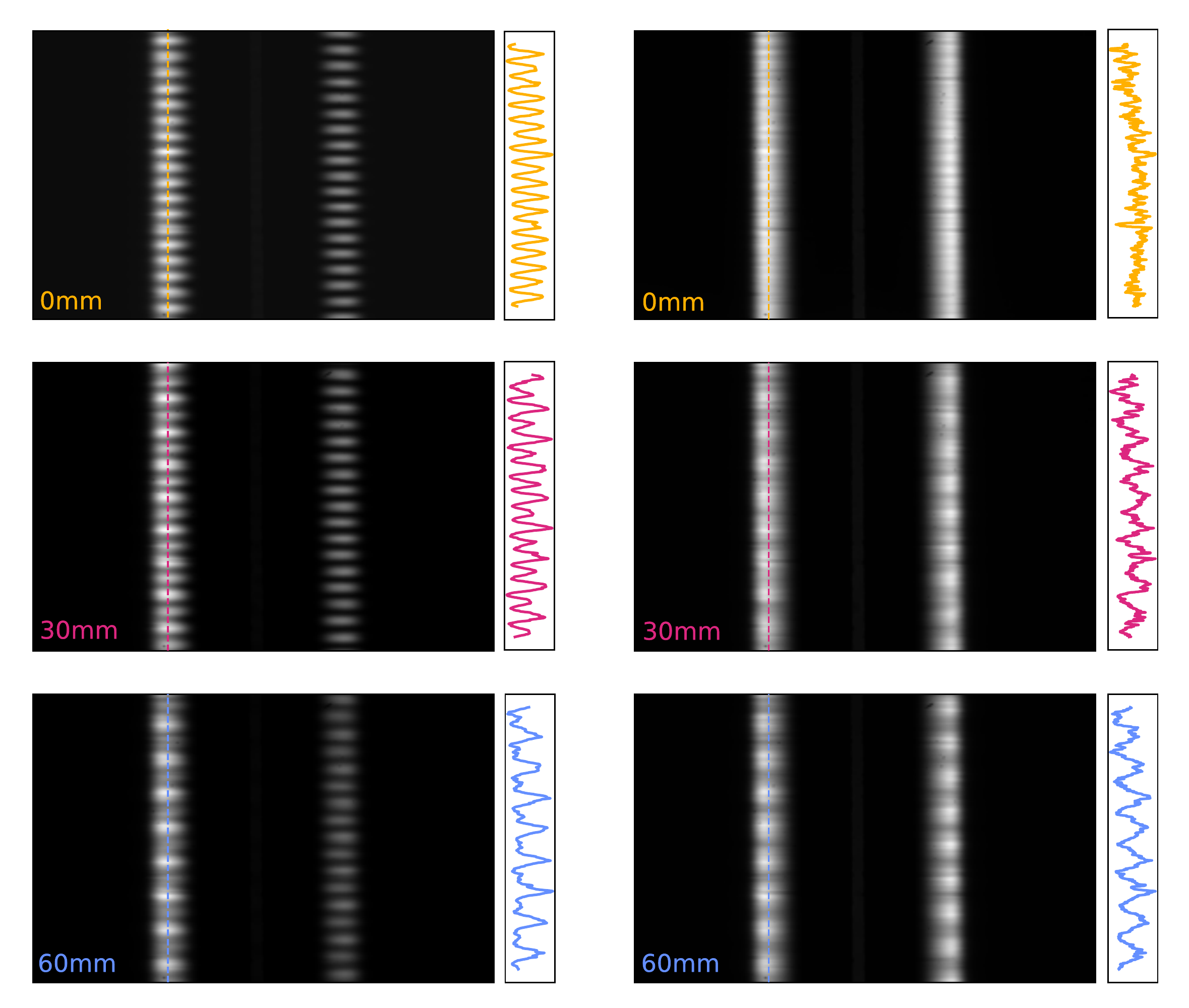}
	\end{tabular}
	\end{center}
   \caption[example] 
   { \label{fig:meas_000030060mm} Q (left column) and unpolarized images (right column) taken with LSDPol$_{\text{QWR}}$. The propagation distances are $1.5 \pm 2.0$mm (referred to as 0mm), $31.5 \pm 2.0$mm (referred to as 30mm) and $61.5 \pm 2.0$mm (referred to as 60mm). The colored lines indicate the location of the cut to the right of the images.  
}
   \end{figure}

It is expected that the intensity modulation resulting from both a polarized and unpolarized light source will change as a function of the propagation distance. Our numerical simulation showed a clear correlation between the two. Figure \ref{fig:meas_000030060mm} displays the \textit{Q} and flat field measurements where the propagation distance was set to 0 mm (top), 30 mm (middle) and 60 mm (bottom). The vertical colored lines in the images represent the locations of the intensity modulation that are plotted next to the images. The images on the right can be seen as flat fields, considering that the measurements were made with a diffuse, unpolarized light source that isotropically illuminates the aperture.

From a quick visual inspection, we conclude that the alternating variation effect found in our simulation, is also present in the lab measurements. As the propagation distance increases, the diffraction effects become more prominent. The images prove that diffraction effects are present in the flat field as well. This effect causes an apparent \textit{V} as it has a frequency that is half the frequency of the Q modulation frequency. It can only be distinguished from a V signal because of its phase shift. This effect is (partly) causing the shifted V lookalike when demodulating any full-Stokes measurement.

\subsection{Influence of increasing distance on the zeroth diffraction order}
\label{sec:meas_0th}
\begin{figure} [ht]
   \begin{center}
   \begin{tabular}{c} 
  \includegraphics[width=0.95\textwidth]{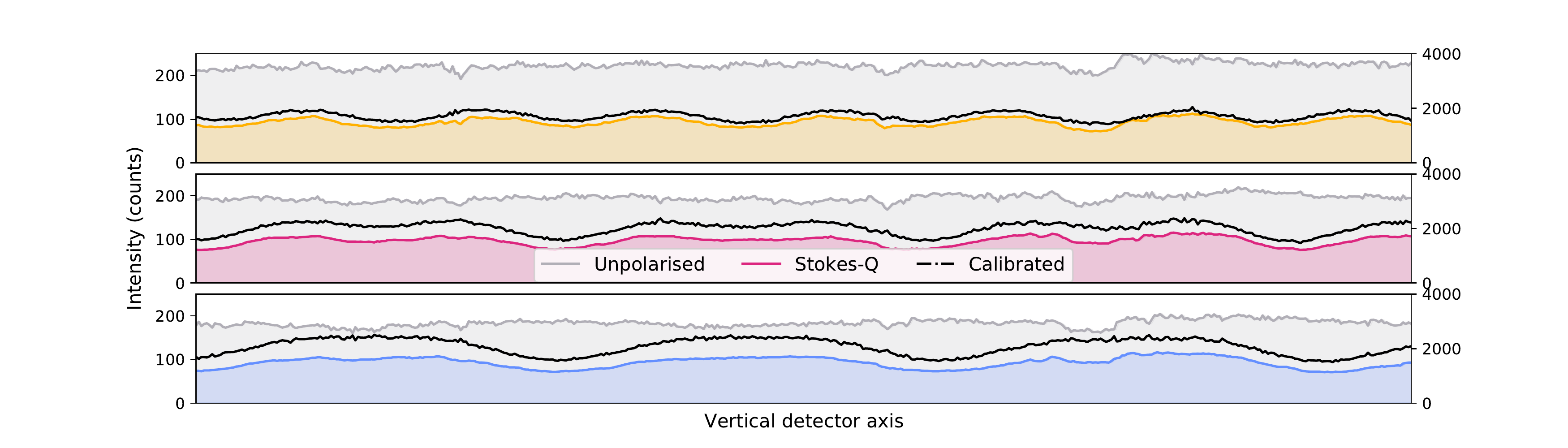}
	\end{tabular}
	\end{center}
   \caption[example] 
   { \label{fig:meas_distance_0th_counts} Intensity modulation present in the zeroth order of \textit{Q} (colors), flat/unpolarized (grey) and calibrated (black dotted) images. The propagation distances are 0 mm(top), 30 mm(middle) and 60 mm(bottom). 
}
   \end{figure}

Inspection of the first diffraction order is done by selecting the central 40 pixel columns of the detector images (Figure \ref{fig:meas_000030060mm}). The median over 40 pixels of the \textit{Q} and flat field is presented in Figure \ref{fig:meas_distance_0th_counts}. A flat-calibrated \textit{Q} response illustrates the importance of a proper flat-field calibration. 

The calibrated \textit{Q} signals in the zeroth order reveal the same frequencies as the dark-subtracted \textit{Q} signals. This is due to the almost homogeneous, flat signal. The modulation frequency of \textit{Q} seems to decrease as a function of distance. The lower graph, corresponding to a distance of 60 mm, reveals a shifted V lookalike.     

\subsection{Influence of increasing distance on the first diffraction orders}
\label{sec:meas_1st}
Inspection of the first diffraction orders is done similar as for the zeroth order. Instead of calculating the median of 40 columns, we simply select two single pixel columns on opposite sides of the detector images.

  \begin{figure} [ht]
   \begin{center}
   \begin{tabular}{c} 
  \includegraphics[width=0.95\textwidth]{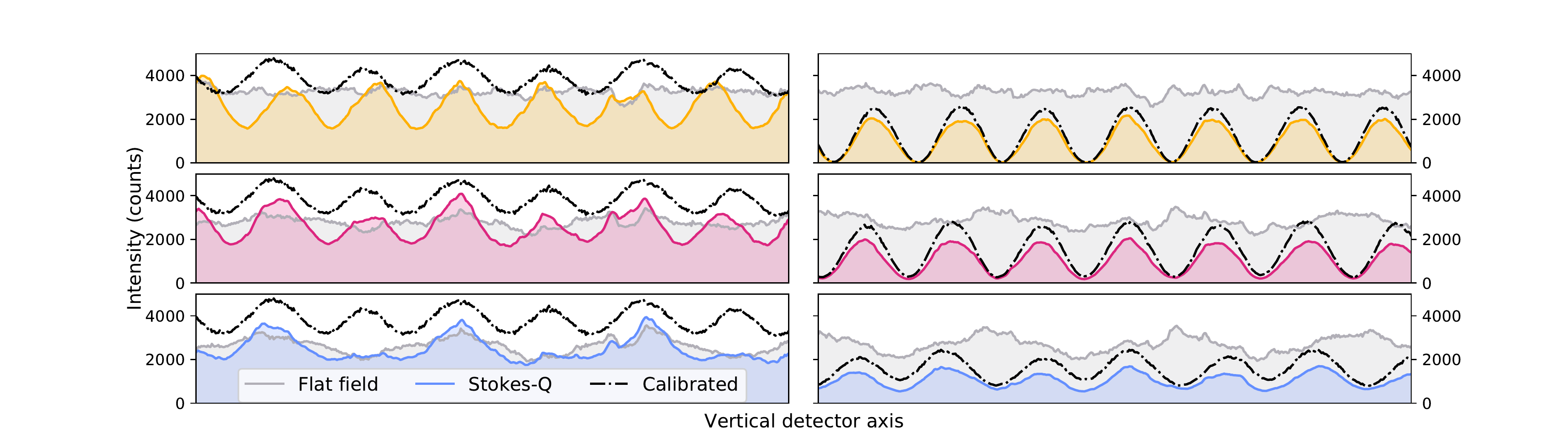}
	\end{tabular}
	\end{center}
   \caption[example] 
   { \label{fig:meas_distance_1st_withcal} The left (m=-1) and right plot (m=1) reveal the intensity modulation present in the first diffraction orders. The calibrated data (black dotted line, identical calibration steps as for the previous Figure) is plotted to compare the raw and calibrated Q. The right axis indicates the intensity in counts of the calibrated signal. 
}
   \end{figure} 
 
Just as for the zeroth grating order, a dependence on the propagation distance can also be seen. The m=1 order is more affected than the m=-1 order, revealing a stronger alternating modulation in m=-1. Looking at the amplitude of the flat, a V signal appears with a phase difference of 90 degrees. When we calibrate the Q with the flat, the flat corrects for part of the alternating modulation in the m=-1 order. The importance of a correct flat is crucial if we can use the flat as a possible diffraction correction. 

\section{Conclusions}
\label{sec:conclusions}
Keller et al. (2020) found an additional spurious ghost signal and a phase-shifted spurious modulation with a frequency equal to the V modulation while preforming data analysis for the LSDPol prototype. Possible explanations that were provided were circular to linear cross talk and diffraction effects of the spatial polarization modulator. 

Our simulations revealed that the two signals can be explained by Fresnel diffraction. The results of these simulations were validated with LSDPol measurements in the lab. 

\begin{itemize}
    \item Fresnel diffraction after the polarization modulator creates an unwanted signal on top of the expected polarization modulation. The size of the effect differs for the zeroth and first grating orders and depends on the polarization of the light source.
    \item The following spurious signals have been identified by increasing the distance between the polarization modulator and the entrance slit:
    \begin{enumerate}
        \item The expected homogeneous zeroth order signal is contaminated by an additional oscillation with a frequency equal to $2\theta$.
        \item The first grating orders reveal an additional modulation on top of the expected intensity modulation. For larger distances, a second oscillation appears on top of the first effect. 
        \item Unpolarized light produces a weak V like modulation that is phase shifted by $\pi/2$ for larger propagation distances. 
    \end{enumerate}
    \item Lab measurements confirm the dependence of the amplitude of the spurious signals on the distance between the polarization modulator and the entrance slit.
    \item The procedure of taking a flat field is crucial. A piece of white office paper is used to create a diffuse, isotropic, white calibration light source. We found that the flat field changes for different types of paper, as well as for a change in the distance between the aperture of the instrument and the paper. 
    \item The flat fields are an important correction factor for an additional shifted  V lookalike. The signal could mimic a linear to circular cross-talk if it is not corrected for properly. 
    \item For the improvement of the LSDPol prototype, it is crucial to understand and take into account the effect of the Fresnel propagation. A full analytical diffraction analysis should explain the detected amplitude change as a function of wavelength that depends on the Stokes vector of the incoming light as well as the grating diffraction order.
\end{itemize}

\acknowledgments 
We want to thank Emiel Por personally for his help regarding setting up the simulations. 
This research made use of HCIPy\cite{Por18}, an open-source object-oriented framework written in Python for performing end-to-end simulations of high-contrast imaging instruments. 

\bibliography{main} 
\bibliographystyle{spiebib} 

\end{document}